\documentclass[sigconf,nonacm]{acmart}

\usepackage{pifont}
\usepackage{arydshln}
\usepackage{amsmath,amsfonts}
\usepackage{algorithmic}
\usepackage{url}
\usepackage{pifont}
\usepackage{arydshln}
\usepackage{graphicx}
\usepackage{textcomp}
\usepackage{preprint_header}
\usepackage{xcolor}
\AtBeginDocument{%
  }

\setcopyright{acmcopyright}
\copyrightyear{2023}
\acmYear{2023}
\acmDOI{XXXXXXX.XXXXXXX}

\acmConference[JCDL '23]{ACM/IEEE Joint Conference on Digital Libraries}{June 26--30,
  2023}{Santa Fe, NM}
\acmPrice{15.00}
\acmISBN{978-1-4503-XXXX-X/23/06}

\begin{document}
\title{TEIMMA: The First Content Reuse Annotator for Text, Images, and Math}

\author{Ankit Satpute}
\email{Ankit.Satpute@fiz-karlsruhe.de}
\orcid{0000-0003-3219-026X}
\affiliation{
  \institution{FIZ Karlsruhe Leibniz Institute for Information Infrastructure}
  \streetaddress{Franklinstrasse 11}
  \city{Berlin}
  \country{Germany}
  \postcode{10587}
}
\additionalaffiliation{
  \institution{George August University of Göttingen}
  \streetaddress{Wilhelmsplatz 1}
  \city{Göttingen}
  \country{Germany}
  \postcode{37073}
}

\author{Andr\'{e} Greiner-Petter}
\email{Greiner-Petter@gipplab.org}
\orcid{0000-0002-5828-5497}
\affiliation{
  \institution{George August University of Göttingen}
  \streetaddress{Wilhelmsplatz 1}
  \city{Göttingen}
  \country{Germany}
  \postcode{37073}
}

\author{Moritz Schubotz}
\email{Moritz.Schubotz@fiz-karlsruhe.de}
\orcid{0000-0001-7141-4997}
\affiliation{
  \institution{FIZ Karlsruhe Leibniz Institute for Information Infrastructure}
  \streetaddress{Franklinstrasse 11}
  \city{Berlin}
  \country{Germany}
  \postcode{10587}
}

\author{Norman Meuschke}
\email{meuschke@uni-goettingen.de}
\orcid{0000-0003-4648-8198}
\affiliation{
  \institution{George August University of Göttingen}
  \streetaddress{Wilhelmsplatz 1}
  \city{Göttingen}
  \country{Germany}
  \postcode{37073}
}

\author{Akiko Aizawa}
\email{aizawa@nii.ac.jp}
\orcid{0000-0001-6544-5076}
\affiliation{
  \institution{National Institute of Informatics}
  \streetaddress{2 Chome-1-2 Hitotsubashi}
  \city{Tokyo}
  \country{Japan}
  \postcode{101-0003}
}

\author{Olaf Teschke}
\email{Olaf.Teschke@fiz-karlsruhe.de}
\affiliation{
  \institution{FIZ Karlsruhe Leibniz Institute for Information Infrastructure}
  \streetaddress{Franklinstrasse 11}
  \city{Berlin}
  \country{Germany}
  \postcode{10587}
}

\author{Bela Gipp}
\email{gipp@uni-goettingen.de}
\orcid{0000-0001-6522-3019}
\affiliation{
  \institution{George August University of Göttingen}
  \streetaddress{Wilhelmsplatz 1}
  \city{Göttingen}
  \country{Germany}
  \postcode{37073}
}

\renewcommand{\shortauthors}{Satpute et al.}
\acmArticleType{Research}
\acmCodeLink{https://github.com/borisveytsman/acmart}
\acmDataLink{htps://zenodo.org/link}
\begin{abstract}
This demo paper presents the first tool to annotate the reuse of text, images, and mathematical formulae in a document pair---TEIMMA.
Annotating content reuse is particularly useful to develop plagiarism detection algorithms.  
Real-world content reuse is often obfuscated, which makes it challenging to identify such cases.
TEIMMA allows entering the obfuscation type to enable novel classifications for confirmed cases of plagiarism. 
It enables recording different reuse types for text, images, and mathematical formulae in HTML and supports users by visualizing the content reuse in a document pair using similarity detection methods for text and math.
\end{abstract}



\maketitle
\thispagestyle{preprintbox}

\section{Introduction}
The effectiveness of an algorithm, particularly in Machine Learning, heavily depends on the quality of the dataset used to develop the algorithm.
Accurate annotations of reused content in a document pair are crucial for developing systems to detect plagiarism, paraphrases, and summaries~\cite{TextReuse}.
Existing plagiarism detection systems (PDS) can only identify copied and slightly altered text~\cite{LitRevPD}.
Developing advanced PDS capable of identifying strongly disguised cases of content reuse requires compiling a gold-standard dataset.
To the best of our knowledge, no tool exists for annotating such cases of content reuse, which is required for creating a suitable dataset.
A few tools allow annotating by selecting text from a single PDF, but none provide functionality for annotating PDF pairs~\cite{pdfAnnowithla}. 
This approach encounters issues such as varying encodings and text representation formats, incorrect character offsets, and undetected non-textual elements~\cite{problEnco}.

Bast et al. have shown the challenges of extracting text from PDFs and the shortcomings of tools supporting this task~\cite{benchMar}.
In particular, scientific documents in the STEM fields (Science, Technology, Engineering, Mathematics) often contain non-textual elements such as mathematical formulae, which are typically ignored during content annotation. 
Extracting text is easier than extracting mathematical expressions because the formula as presented in a PDF does not allow capturing the formula's structure or semantics, available in LaTeX or MathML\footnote{\url{https://www.w3.org/Math/}}~\cite{zbMATHtoLatex}.
Annotating math is a highly complex task supported by specialized tools to enrich mathematical formulae, such as MioGatto~\cite{mioGatto}, MathAlign~\cite{mthAlign}, and AnnoMathTeX~\cite{annomathTex}.
These tools allow to save math in its original form, such as LaTeX or MathML, but none support recording annotation on a document pair.

This paper introduces TEIMMA, the first tool that enables the annotation of reused text, images, and math in their original transcribed form. TEIMMA's \textbf{source code} is publicly available~\cite{ourGitHub}, and we provide a \textbf{live demonstration} of TEIMMA's features at \texttt{\url{https://teimma.gipplab.org}}.

\begin{figure*}
\includegraphics[width=1\textwidth]{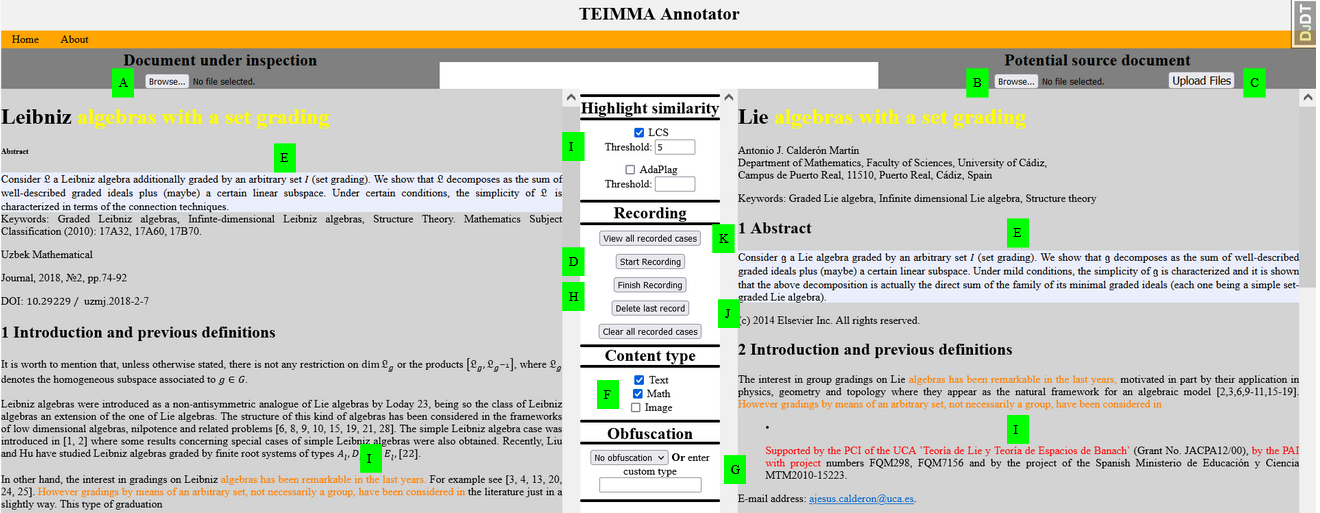}
\vspace{-6mm}
\caption{TEIMMA User interface. The left document~\cite{zbMATH07374976} has been retracted for plagiarizing the right document~\cite{zbMATH06300457}.} 
\label{Userinterface}
\end{figure*}

\section{Architecture and Use}\label{subsec:systemoverusage}
TEIMMA (TExt, IMage, MAth) is a web-based tool to visualize and annotate similar content in document pairs using a machine-processable format.
We refer to similar formulae, text, or a combination thereof as a case.
The tool stores annotated cases of similar text as plain text, similar math as MathML, and similar images as IDs referring to the original images.
Documents previously annotated with TEIMMA can be re-uploaded to modify and add annotations.

The tool accepts documents in PDF, LaTeX, and plain text (.txt)\footnote{Files in .txt format do not support image or math annotations} format.
TEIMMA performs a multi-step conversion of PDFs to HTML and MathML to ensure the accurate representation of text, images, and math. 
First, it uses the open-source Python package \textit{pdftolatex}~\cite{kanigicherla} to extract the positions of text, math, and images from the PDF.
The package converts text to LaTeX and math to images.
Second, TEIMMA employs the LaTeX OCR model \textit{pix2tex}~\cite{githubGitHubLukasblecherLaTeXOCR} to convert the images of math formulae returned by \textit{pdftolatex} to LaTeX.
The model uses a pre-trained Vision Transformer encoder~\cite{2020vitencoder} with a ResNet backbone and a Transformer decoder trained on the im2latex-100k dataset~\cite{2017transformers,im2latex}. 
In the third step, TEIMMA combines the extracted text, images, and math to create a complete LaTeX representation of the PDF.
If possible, we recommend using input documents in LaTeX because PDF to LaTeX, like any other conversion, entails the risk of errors.  
Thus far, no comprehensive evaluation of the conversion accuracy for math extraction from PDF exists \cite{wang2020pdf2latex,sur2023mathematical,MeuschkeJSM23}.
Lastly, TEIMMA converts the LaTeX output to HTML and MathML for mathematical content using LaTeXML~\cite{nistLATExmlLATEX}---the best-performing tool for this task~\cite{Schubotz2018}. 

TEIMMA uses HTML tag names to extract text, images, and math~\cite{parallel} and records annotations in terms of the character positions of selected content in a plain text file.
The tool replaces each math formula in MathML format with its assigned ID while maintaining its start character position in the plain text file.
This allows separating formulae from the plain text to prevent the typically extensive MathML markup of formulae from distorting the character positions.

Figure~\ref{Userinterface} shows TEIMMA's user interface for visualizing and annotating similar content.
The buttons \hyperref[Userinterface]{\colorbox{green}{A}}, \hyperref[Userinterface]{\colorbox{green}{B}}, and \hyperref[Userinterface]{\colorbox{green}{C}} allow uploading the two documents for investigation. 
TEIMMA converts both documents to HTML and saves the extracted plain text, math formulae, and images in the database.
After clicking the \textit{Start Recording} button \hyperref[Userinterface]{\colorbox{green}{D}}, users select a span in both documents. 
TEIMMA extracts the text from the span and matches it to the text in the plain text file to obtain the span's start and end character positions.
The selected span is highlighted by assigning it a unique background color \hyperref[Userinterface]{\colorbox{green}{E}}. 
The checkboxes in the \textit{Content type} section \hyperref[Userinterface]{\colorbox{green}{F}} allow configuring the type of annotations to be performed, e.g., only annotations of similar text.
The section \textit{Obfuscation} \hyperref[Userinterface]{\colorbox{green}{G}} allows users to enter the obfuscation type, e.g., paraphrase or summary, they think has been used to obfuscate the content.
Users can activate one of four algorithms\footnote{Note: Only two algorithms are visible in Figure~\ref{Userinterface} due to space limitations.} \hyperref[Userinterface]{\colorbox{green}{I}} to receive support with annotating by viewing similar text and math content in the uploaded document pair.
To view similar text, users can choose between the longest common substring (LCS) or AdaPlag, the winning method in the latest PAN plagiarism competition~\cite{sanchez2015adaptive}.
For similar math tokens, the longest common identifier sequence (LCIS) or greedy identifier tiling (GIT)~\cite{normanCitplag,Meuschke21}.
Moreover, users can specify the minimum length required for displaying the matches that each algorithm identified. 
For the text-based algorithms LCS and AdaPlag, the length threshold represents the number of words, and for the math-based algorithms LCIS and GIT, the number of math symbols in the match.
The \textit{Finish Recording} \hyperref[Userinterface]{\colorbox{green}{H}} button saves the recorded span in the database along with the data entered by users for describing the identified similarity. 
The \textit{Delete the last record} \hyperref[Userinterface]{\colorbox{green}{J}} button deletes the previously recorded annotation. 
Saved annotations in the database can be viewed and downloaded as a JSON Lines (.jsonl) file by clicking on \textit{View all recorded cases} \hyperref[Userinterface]{\colorbox{green}{K}} button.
Users must create annotations for each document pair separately if a document shares content with multiple other documents. 
However, TEIMMA keeps track of overlapping annotations by checking if previous annotations for re-uploaded documents exist in the database.

The final annotation stored in the database in JSON format contains document names, the character offsets for the start and end of text spans, and the IDs for images and formulae.
The original images and formulae in MathML are also stored in the database.
Additionally, the annotation contains the content and obfuscation type if users entered them.



\section*{Acknowledgment}

This work was funded by the Deutsche Forschungsgemeinschaft (DFG, German Research Foundation) – 437179652, the German Academic Exchange Service (DAAD) - 57515245, and the Lower Saxony Ministry of Science and Culture and the VW Foundation.

\bibliographystyle{ACM-Reference-Format}
\bibliography{_references}

\end{document}